\documentclass[aps,prd,numbers,
preprint,
floats,amsmath,amssymb,amsfonts,groupedaddress,
showpacs,showkeys
 a4paper,twoside
 ,byrevtex]{revtex4-1}

\usepackage{bm}
\usepackage{graphicx}
\usepackage{graphics}
\usepackage{float}
\usepackage{latexsym}
\usepackage{epsfig}
\usepackage{dcolumn}
\usepackage{url}

\usepackage{lmodern} 
\usepackage[T1]{fontenc}
\usepackage{pdfsync}
\usepackage[pdftex,bookmarks,colorlinks,breaklinks]{hyperref}
\hypersetup{linkcolor=black,
citecolor=blue,filecolor=red}

\newcommand{\Hcal}{{\mathcal H}}

\newcommand{\Cc}{{\mathbb C}}

\newcommand{\half}{{\textstyle \frac{1}{2}}}

\newcommand{\bvec}{{\bf b}}

\newcommand{\svec}{{\bf s}}

\newcommand{\evec}{{\bf e}}

\usepackage{epstopdf}

\begin{document}
\title{Cosmic Forms}

 \author{Maurice Kleman}\email{kleman@ipgp.fr}

\affiliation{\vspace{1cm}Institut de Physique du Globe de Paris $-$ 
 Sorbonne Paris Cité\\
 1, rue Jussieu - 75238 Paris c\'edex 05\\}

\begin{abstract}

The continuous 1D defects of an isotropic homogeneous material in an Euclidean 3D space are classified by a construction method, the Volterra process (VP). We employ the same method to classify the continuous 2D defects  of a vacuum in a 4D maximally symmetric spacetime. These defects, which we call \textit{cosmic forms}, fall into three different classes: i)- $m$-forms, akin to 3D space disclinations, related to ordinary rotations and analogous to Kibble's global cosmic strings (except that being continuous any deficit angle is allowed); ii)- $t$-forms, related to Lorentz boosts (hyperbolic rotations); iii)- $r$-forms, never been considered so far, related to null rotations. A detailed account of their metrics is presented. In each class, one distinguishes between wedge forms, whose singularities occupy a 2D world sheet, and twist or mixed forms, whose inner structure appears as a non-singular \textit{core} separated from the outer part by a 3D world shell  with distributional curvature and/or torsion. Relaxation processes of the world shell involve new types of topological interactions between cosmic dislocations and cosmic disclinations. The resulting structures of the core region itself, whether wedge or not, are not explored in detail in this article.
 
Whereas $m$-forms are \textit{compatible} with the (narrow) cosmological principle (nCP) of space homogeneity and isotropy, $t$- and $r$-forms demand spacetime homogeneity (perfect CP).
 Thus we advance that $t$- and $r$-forms are typical of a primeval false vacuum obeying the perfect CP, inhabiting a de Sitter spacetime. 
 
Cosmic forms may assemble into networks with conservation laws at their nodes, such that all the segments are made of \textit{positive} forms, say, thus generating some characteristic curvature field. To this network may be adjunct a conjugated network made of \textit{negative} forms, in order to tune the final spatial curvature to a given value.

\end{abstract}

\date{\today}

 

\maketitle
\normalsize

\section{Introduction} \label{int}
The concept of macroscopic \textit{cosmic strings} has been investigated to a considerable extent in the last three decades, starting from the pioneering work of \textcite{kibble76}; see \cite{vilenkin94,*hindmarsh95} for reviews. According to these theories, cosmic strings originate in particle phase transitions, like the grand unification ($t \approx 10^{-36}$ s, before the inflationary epoch) or the electroweak transition ($t \approx 10^{-10}$ s, long after the inflationary epoch). 
 In the inflation picture of the universe, these transitions are also described as occurring between \textit{vacua} of decreasing energy density, the higher energy vacua being qualified of \textit{false vacua} and the present one, in which we are living, of \textit{true vacuum}. Cosmic strings are \textit{topological} singularities and as such should have lasted up to the present day. Their existence could in principle be detected by gravitational lensing, see e.g. \cite{gott85}, but in spite of considerable efforts has not yet been assessed. Nevertheless they continue to arouse a great interest, for many reasons: e.g. \textcite{zeldovich80} has argued that such strings produced in the protogalactic universe could provide the fluctuations that account for the formation of galaxies; \textcite{polchinski07} has claimed that these objects are macroscopic versions of cosmic superstrings; their possible role in the small-angle CMB temperature anisotropies are the object of many investigations, cf. e.g., ref. \cite{fraisse08,*durrer10}. Cosmic strings thereby constitute an important feature of particle physics motivated scenarios of the history of the universe.

The topology of a cylindrically symmetric cosmic string is often presented as resulting from a \textit{cut-and-glue process} performed on a Minkowski spacetime $M^4$, as follows: a wedge of angle $\alpha$, whose 2D edge is along a straight singularity (the \textit{world sheet}), here a 2D plane, is cut out from the spacetime along an hyperplane bordered by this world sheet and the two lips of the wedge identified. This is much akin to the familiar \textit{Volterra process} (VP) in Condensed Matter Physics (CMP)\cite{friedeldisloc}, which however presents some possibilities of generalization not present in the usual cosmic string theory. This is the main topic we wish to develop in this paper. We call \textit{cosmic forms} the objects classified in this way.

In analogy with its CMP definition, but with one dimension more, a general VP carried out in a maximally symmetric spacetime would start from the consideration of a 3D manifold of{ any shape}, the so-called \textit{cut hypersurface} $\Sigma$ (a generalization of the hyperplane above), bordered by a 2D manifold designed to be the singular cosmic disclination, i.e. the {world sheet}. A cut is then carried out along $\Sigma$ (hence its name) whose two \textit{lips} are displaced relatively one apart the other under the action of an element $g \in G$ of the group of isometries $G$ of the spacetime (the rotation $\alpha$ is such an isometry), then the sector of spacetime thus obtained is, according to the case, either removed and the lips identified (as in the cut-and-glue process) or filled with an element of spacetime that matches exactly with the parted lips. In CMP the relative displacement of the two lips is qualified of \textit{rigid}, to imply that this relative displacement is a space symmetry element.

The symmetry $g$ that is broken in a cosmic form is continuous, because $G$ is a continuous group. Hence these defects are not topological defects and can smoothly vanish should the necessity arise. If the isometry is a continuous angle $\alpha$ (in the language of differential geometry, a continuous rotational holonomy), it would
 necessarily match for some value to the quantized $\alpha$ resulting from the gauge field theory of cosmic strings. Therefore
the calculation of cosmic strings in the case of a rotation $\alpha$ borrows from the methods of the standard Volterra defect theory, with delta-function valued \textit{curvature} components on the line itself, cf. e.g. \cite{vilenkin81b,*hiscock85,*tod94,puntigam97}; the \textit{core} region of a cosmic string is indeed assumed to be concentrated along the world sheet, in a region whose size is comparable to a Planck length. 

Therefore cosmic forms can be classified, as the foregoing discussion suggests, according to the symmetries of the spacetime, and in particular those of the Lorentz group. 
Dislocations (with continuous Burgers vectors or, in the language of differential geometry, continuous translational holonomies) can as well enter such a classification, if extended to the full spacetime group of isometries, with delta-function valued \textit{torsion} components on the line itself, resulting from the presence of spinning matter. A complete picture of spacetime in terms of continuous defects, involving continuous curvature and continuous torsion, viz. continuous disclinations and continuous dislocations, has been elaborated, in the frame of the Einstein$-$Cartan$-$Sciama$-$Kibble (ECSK) theory of general relativity, cf. e.g. \textcite{hehl} for a review. The analogy with the continuous theory of defects in CMP is noticeable \cite{kroner60,*kroner}.  

With these generalizations we are clearly getting quite far from the realm of macroscopic cosmic strings. Let us recall that usual cosmic strings are restricted to singularities carrying a discrete rotational holonomy $\alpha$, whose world sheet is a 2D plane invariant under the action of this rotation; they are akin to CMP \textit{wedge} disclinations $-$ to the exception of \textcite{puntigam97} who have considered non-wedge cosmic strings. CMP considers indeed objects much more general than wedge disclinations (viz., \textit{twist} and \textit{mixed} disclinations); in these latter cases the disclination line is not along the rotation axis.
 There is no obvious reason not to take seriously these generalizations  as illustrating new physical features of the spacetime. This point of view, if systematically pursued, opens quite novel considerations on the nature and the role of cosmic line defects: 
 
\noindent $-$ on their nature: whereas cosmic strings stem from phase transitions, the cosmic defects we are contemplating originate in the context of forced spacetime curvature or torsion variations, 
 
 \noindent $-$ on their role: cosmic forms can form networks with topological conservation relations at their nodes. If they do, the question arises whether they relate in some way to the large-scale structure of the Universe at the present epoch or at some earlier epoch after which they could have  softly vanished away. \\ \vspace{-10pt}
 
 This paper is built as follows.
 
 We recall in Sect.~\ref{classdef} that there are two categories of defects of the order parameter of an homogeneous medium in CMP (these results extend straightforwardly to spacetimes):

$-$  those that come under the heading of algebraic topology (\textit{topologically stable defects}, which are but to a few exceptions true \textit{singularities} of the 'order parameter'), These defects are related to the homotopy groups of a manifold characterizing this order parameter,
 
$-$  those whose characteristics $-$ Burgers vector, rotation or Frank vector, either continuous or quantized \footnote{In that case these defects are also topologically stable} $-$, are related to the elements of the symmetry group. 
These are called \textit{Volterra defects}, and may be constructed by the so-called Volterra process; they are two-dimensional defects in spacetime, i.e., line defects in a 3D submanifold of the spacetime. 

A somewhat more detailed overview of the theory of defects or singularities in CMP, in what it differs or not from the theory of singularities in cosmology, can be found on the \verb"arXiv" site \cite{kleman09}, to which we refer the reader that is little conversant with these matters. 

 Sect.~\ref{poinforms} presents the classification of generic \textit{cosmic forms} including  \textit{cosmic disclinations} and \textit{cosmic dislocations}.
We first reassess in a new perspective the results of ref. \cite{kleman09}, bearing on the classification of Minkowski spacetime defects in the same theoretical frame as the theory of defects in CMP. This classification relies on the fact that the subgroups of the orthochronous connected Lorentz group $L_0$ can be partitioned into four sets of classes of conjugacy. On this basis, our classification retrieves the two different kinds of cosmic strings described in the literature (\textcite{tod94,puntigam97}), plus a third one. We call: 

$-$ \textsf{$m$-forms} the cosmic forms related to continuous symmetry rotations $\alpha$,

$-$ \textsf{$t$-forms} those related to Lorentz boosts (hyperbolic rotations),

$-$ \textsf{$r$-forms} those related to null rotations; $r$-forms have never been considered so far.  

$-$ the fourth group of conjugacy is the Lorentz group in its entirety, and thus contains all the previous forms and no other one.

Wigner, in an epoch-making article \cite{wigner39}, has demonstrated the equivalence between the unitary irreducible representations of subgroups of $L_0$ and the elementary particles. Thus, the subgroups belonging to the same class of conjugacy correspond to a well-defined type of particles: continuous symmetry rotations are related to massive particles, hyperbolic rotations to tachyons and null rotations to massless particles. We shall use these relationships, and assign the fourth group of conjugacy to an 'amorphous' substance, in the sense that it has the same symmetries as the underlying spacetime. Tentatively, we  identify this substance with a quantum field theory \textit{vacuum}.

In Sect.~\ref{forms} we calculate the line elements $ds^2$ of these three classes, not only for the \textit{wedge forms} (two such line elements have already been well investigated for strings) but also for the \textit{twist forms}. In CMP 'wedge' designates a disclination whose axis of rotation is along the line defect, and 'twist'  a disclination whose axis of rotation is perpendicular to it. This terminology is perfectly acceptable for the $m$-forms, which are topologically equivalent to CMP disclinations in a glass. As we shall see, twist CMP disclination has to be topologically stabilized by dislocations or other disclinations attached to it \cite{klfr08}. In fact this process of attachment explains why closed disinclination loops are observed in liquid crystals; it also plays a crucial role in their dynamics. The same process of topological stabilization has to occur for twist forms in spacetime, and more generally for any non-wedge, {mixed}, segment along a curved $m$-form. Thus the topological equivalence between CMP disclinations and $m$-forms is complete. But for $t$- and $r$-forms the element of symmetry carried by the form is a much more subtle object than a rotation. However curved $t$- and $r$-forms can also be topologically stabilized by attached forms. In all those cases where the presence of attached forms seems a requisite, the singularity of the form is not along the world sheet, but along an hypersurface that surrounds the central world sheet, which we call a \textit{world shell}; the core appears as hollow.

Sect.~\ref{disc}, in a first part, takes stock of these results. The second part is mostly speculative. 
We start from the remark that $t$- and $r$-forms are classified by elements of symmetry (in the sense of VP) that do not belong to the maximally symmetric group that encompasses the \textit{space} properties of isotropy and homogeneity $-$ as such it can be said that they do not obey the narrow cosmological principle (nCP), whereas the $m$-forms are {compatible} with this principle. In fact they are compatible with the perfect cosmological principle (pCP) of  Bondi et al. and of Hoyle \cite{bondi48,*hoyle48}, for which the embedding spacetime is de Sitter $dS_4$. Thus we claim that $r$- and $t$-forms are typical of a false vacuum with a positive cosmological constant, which could be met during the process of inflation \cite{linde05ht}. Also the remarkable hollow nature of the core of twist forms might play a role in the decay of the false vacuum.

Finally we stress that cosmic forms can assemble into simple or dual networks of opposite signs which can adjust to any curvature $-$ as disclinations do in amorphous solids \cite{kleman79,kleman83} or in Frank and Kasper phases \cite{frank58b,*frank59}, where they adjust to zero Euclidean curvature. This might be in direct relation to the curvature of the spacetime at the time of inflation.\vspace{-10pt}

\section{The classification of defects in CMP ordered and disordered media}\label{classdef} \vspace{-10pt}
The concept of defect is rather well demarcated in CMP. 
There are two different approaches \citep{klfr08}, which distinguish {\textit{Volterra defects} in direct relation with the elements of the symmetry group, and \textit{quantized, topological defects}. Volterra defects are either continuous or quantized. Thus some of them belong also to the second category.

This section briefly revisits the standard results of the CMP theory of defects in three dimensions, but the concepts thus put forward apply to any dimension.\vspace{-10pt}
 
\subsection{Volterra continuous defects}\label{PVVP} \vspace{-10pt}

The \textit{Volterra process}, first developed in view to classify the singular solutions of Hooke's elasticity in an isotropic homogeneous solid (i.e. an amorphous material, also called a glass), is a construction method of these singularities \citep{friedeldisloc}. We have above presented the VP for a rotational symmetry, but the same process can be adapted to any element of the symmetry group of an ordered or a disordered medium. Let $\Sigma$ be a hypersurface drawn in a unstrained \textit{amorphous }specimen, bordered by a closed or an infinite line L$=\partial \Sigma$; the sample is then cut along $\Sigma$, and the two lips $\Sigma^+$ and $\Sigma^-$ of this \textit{cut surface} $\Sigma$ are displaced relatively one to the other by a \textit{rigid} displacement ${\textbf{d}(\textbf{x})}=\textbf{b}+\bm \Omega(\textbf{x})$; $\textbf{b}$ is a translation and $\bm \Omega$ a rotation, $\textbf{x}$ is any point on $\Sigma$. 
After addition of missing material in the void (if a void is created) or removal of superfluous matter (if there is multiple covering), the atomic bonds are reset on the lips and the specimen let to elastically relax. 
As it can be shown, at the end of this process the singularities of the strain field are restricted to the line itself and the position of the lips is unmarked. 
\begin{figure}[b]
\includegraphics[width=2.5 in]{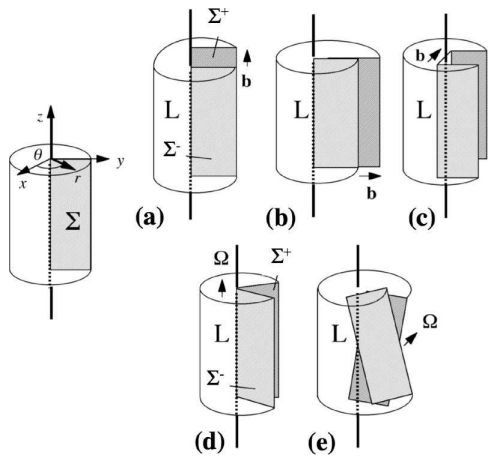}
   \caption{Schematic representation of the Volterra process for a straight line that borders a half plane cut surface $\Sigma$. (a) screw dislocation; (b, c) edge dislocations: these two processes that relate to the same dislocation (but to a global $\mathrm{\pi}/2$ rotation for $\textbf{b}$) yield the same dislocation in an isotropic medium, after relaxation of the stresses; (d) wedge disclination; (e) twist disclination. \textit{adapted from} \citet{klfr08}.}
 \label{fig01} 
\end{figure}

A line defect that carries a translational symmetry $\textbf{b}$, is called a \textit{dislocation}, $\textbf{b}$ is its \textit{Burgers vector}; a line defect that carries a rotational symmetry $\bm \Omega$ is a \textit{disclination}, $\bm \Omega$ is its \textit{rotation vector} (or \textit{Frank vector}). 

Notice that $\textbf{b}$ and $\bm \Omega(\textbf{x})$ are elements of symmetry of the glass, and also of $E^3$, its \textit{habit space}. The group of symmetry $E(3)\sim SO(3)\ltimes\mathbb{R}^3$ of an euclidean manifold $E^3$ being continuous, the Volterra defects are \textit{continuous}, in a sense that will be made more precise later on.

Fig.~\ref{fig01} illustrates various possible dislocation or disclination lines in function of the orientation of the Burgers or rotation vectors with respect to the line L, assumed here to be an infinite straight line. Fig.~\ref{fig01}(c),(d) illustrate the creation of voids, hence addition of matter during the VP. The opposite (removal of matter) can be easily imagined. This choice can be correlated to an \textit{orientation} given to the line according to some arbitrary rule. Fig.~\ref{fig01}(a),(b),(e) also have their opposite construction, so that a defect line is an oriented line in all cases.  \vspace{-10pt}

\subsection{Volterra topological defects}\label{PVVP} \vspace{-10pt}

The Volterra process can be extended to an \textit{ordered} medium, provided $\textbf{d}$ is an element of symmetry of the system. 
In that case the line defects are \textit{quantized} if $ H$ (the {Bravais group}) is discrete, as in a crystal. Because these defects cannot vanish continuously, they are also called topological.
Since $\textbf{d}$ is an element of symmetry $\in H$, there is no singularity of the order parameter along the lips of the cut surface, after completion of the process \citep{friedeldisloc}. For the same reason the cut surface is unmarked after the completion of the VP in the case of a glass.

In ordered media like crystals or liquid crystals the elements of the symmetry group $H $ correspond to possible operations of symmetry in space ($H$ is a subgroup of the space group $G=E(3)$). However line defects can also be considered in other types of media where the Volterra process does not make sense, e.g. vortex lines in superconductors and superfluids: the group of symmetry is then a \textit{gauge group} $H$, which itself breaks some gauge symmetry group $G \nsim  E(3)$. $H$ as a gauge group is not a subgroup of $E(3)$, as a rule. Therefore their relation to a Volterra process is somewhat artificial.}
But there is a \textit{unique} theoretical frame in which quantized Volterra and gauge line singularities both appear as defects, as follows.  \vspace{-10pt}

\subsection{Topological defects}\label{topdefa} \vspace{-10pt}
Let $H \subset G$ a subgroup of $G$, $x \in M$ an element in a space $M$ on which $G$ is acting. The elements of $G$ under which $x$ is invariant form a subgroup $H \subset G$, also called the \textit{little group} or the \textit{isotropy group}; all the elements $\big\{g\,.\,x|\, g\,.\,x \in M, g \in G\big\}$ of its orbit 
are invariant under the subgroups $g\, H \, {g} ^{-1}$ conjugated to $H$. They form a conjugacy class $\left[G:H \right]$. The topological defects of $M$ 
are classified by the non-trivial elements of the homotopy groups $\pi_{i}(\cal{M})$ of the manifold $\mathcal{M}=G/H$, which is the set of the left cosets $gH\,(g \in G)$ of $H$; 
$\cal{M}$ is the \textit{coset space}, also called the degenerate vacuum manifold or the order parameter space \citep{toulouse76}. There are therefore as many coset spaces attached to a group $G$ as there are conjugacy classes $\left[G:H \right]$. 
The topological classification of defects extends to any physical medium with a symmetry group $G$ broken to $H$. $H$ is either the {symmetry group} of the crystal, or a group encompassing more subtle symmetries, like magnetic or electric properties, or a {gauge group}. 

The topological approach extends to defects of any dimensionality (wall defects are classified by $\pi _{0} (\cal{M})$, line defects by $\pi _{1} (\cal{M})$, point defects by $\pi _{2} (\cal{M})$, configurations by $\pi_{3}(\cal{M})$). For reviews see \cite{mermin,*michel,*kleman82b,*trebin}. 

The Volterra approach relates to the case when $H$ is the group of symmetry of some ordered medium whose habit space is $M$; $H$ is also a subgroup of a group $G$ under which $M$ is invariant. It investigates not only the quantized line defects belonging to the non-trivial classes of the fundamental group $\pi_{1}(\cal{M})$ (i.e., topologically stable), but also the continuous defects alluded to previously and that can be topologically assigned to the trivial class $\{1\}\in \pi_{1}(\cal{M})$ (not topologically stable). Volterra continuous defects play an important role in CMP, in particular in liquid crystals \cite{klfr08} where a physically significant part of the symmetry groups is continuous. Cosmic forms are defects of that kind.

Let us recall that the analogies between Volterra quantized defects in liquid crystals and cosmic strings have inspired 'cosmology in the laboratory' experiments, see \cite{chuang91,*digal99,*bowick94}. \vspace{-15pt}

 \section{Forms in a Minkowski spacetime $M^4$} \label{poinforms} \vspace{-10pt}

\subsection{The classification of cosmic forms} \label{cosmicforms} \vspace{-10pt}

We restrict our discussion to the Minkowski spacetime $M^4$; the extension to a de Sitter space is straightforward, see Sect.~\ref{forms-dislo}. 

\begin{figure}[h]
\includegraphics[width=2. in]{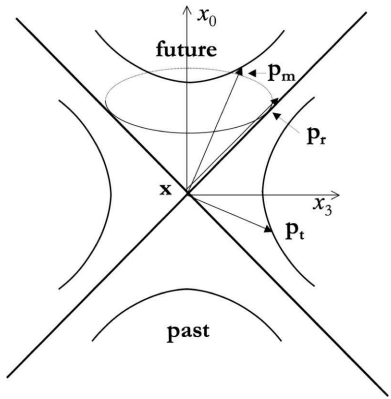}
   \caption{ 3D representation of the light cone at some event $\textbf{x}$, cut in the $\mathbf{e}_0\,\mathbf{e}_3$ plane. 
   The three non-trivial orbits of the extremities of the vectors ${p}^a_m$ (timelike), ${p}^a_r$ (null), ${p}^a_t$ (spacelike) are respectively hyperboloids of two sheets, the light cone with the origin excluded, hyperboloids of one sheet. A cross section of the future light cone is sketched in perspective; it is a 2-sphere in 4D. The origin ${p}^a=0$ is a 4$^{rth}$ orbit in itself.}
 \label{fig02} 
\end{figure}

The group of isometries $P(4) \sim L_0\ltimes \mathbb{R}^{1,3}$ of $M^4$ is partitioned into four conjugacy classes, yielding four corresponding coset spaces. \textcite{michel} obtains these conjugacy classes by considering the action of  $P(4)$ on 4-momenta $p^a$, each orbit corresponding to a different norm of $\|{p^a}\|^2$, each stratum to a set of orbits corresponding respectively to timelike, spacelike, null, and vanishing momenta, Fig.~\ref{fig02}. The use of the group action on 4-momenta for this calculation is somewhat arbitrary, but can be given a cosmological interpretation, at least for timelike and null momenta, within the scope of Wigner's theory \cite{wigner39}, see Sect.~\ref{stab-cryst}.\\ \vspace{-10pt}

 \textsf{1- Timelike 4-momenta} 
 
    Let ${p}_m^a$ be some timelike 4-momentum. The little group $H_m$ that leaves ${p}_m^a$ invariant is isomorphic to the group of 3D rotations; $H_m\sim {SO(3)}$. 
    The orbit of ${p}_m^a$ (under the action of all the elements of $L_0$) belongs to a hyperboloid of two sheets. In Fig.~\ref{fig02}, where we have taken $p^{0}>0$, i.e. it is the sheet belonging to the future of the event $\textbf{x}$. 
    
        The coset space $\mathcal{M}_m =L_0/SO(3)$ is topologically equivalent to the orbit, i.e., $\mathcal{M}_m=H^3$. 
The corresponding homotopy groups are all trivial
\begin{equation}
\label{e1}
  \pi_{i}({H^3)\sim \{1\}}.
\end{equation} 

We are left with continuous Volterra defects, \textit{$m$-dislocations} and \textit{$m$-disclinations}, characterized by the elements of the full little group of symmetry, translations and rotations included, $ P_{m} \sim H_m\ltimes \mathbb{R}^{1,3} \sim SO(3)\ltimes\mathbb{R}^{1,3}$. \\ \vspace{-10pt}
 
 \textsf{2- Spacelike momenta} 

The little group $H_t$ that leaves a spacelike 4-momentum ${p}^a_t$ invariant is isomorphic to ${PSl(2,\mathbb{R}) \sim Sl(2,\mathbb{R})/\mathbb{Z}
_2}$ \cite{sternberg94}. These are Lorentz boosts. The coset space $\mathcal{M}_t$ is topologically equivalent to a spacelike hyperboloid of one sheet, i.e. $\mathcal{M}_t\sim S^2 \times \mathbb{R}$; thus:
 \begin{equation}
\label{e3}
  \pi_{i}(\mathcal{M}_{t})\sim \pi_{i}(S_{2}) \rightarrow \\ 
   \pi_{1}(\mathcal{M}_{t})\sim \{1\},\,\pi_{2}(\mathcal{M}_{t})\sim \mathbb{Z},\,\pi_{3}(\mathcal{M}_{t})\sim \mathbb{Z},\,\pi_{4}(\mathcal{M}_{t})\sim \mathbb{Z}_2.
\end{equation}
Hence there are both Volterra continuous and topologically stable defects, whose dimensionalities in $M^4$ are: 1 for $\pi_{2}(\mathcal{M}_{r})$ and 0 for $\pi_{3}(\mathcal{M}_{r})$. $\pi_{4}(\mathcal{M}_{r}) $ designates possible \textit{configurations} (in the sense of \textcite{michel}).
All line defects are continuous defects. By definition, they obtain by a VP performed on the full spacetime, defined by any element of the subgroup $P_{t}\sim H_t\ltimes \mathbb{R}^{1,3}$. Notice that these lines defects appear here as the continuous defects of a 2D hyperbolic plane or any homotopic manifold, but there is no such manifold of easy physical significance in the spacetime; it is therefore more sensible to attach the $H_t$ Volterra forms to the full group of isometry of the spacetime.
 \\ \vspace{-10pt}

     \textsf{3- Null momenta} 
     
     The little group $H_r$ that leaves ${p}_r^a$ invariant is isomorphic to the 2D Euclidean group; $H_r\sim {E}(2)= SO(2)\ltimes\mathbb{R}^2 $, the semidirect product of a circle by a plane.

    The (unique) orbit of ${p}_r^a$, which is also the coset space $\mathcal{M}_r = L_0/E(2)$, is  topologically equivalent to the future null cone with apex removed, $ \|{p}_r^a\|^2=0$, i.e., the cartesian product $\mathcal{M}_r\sim S^2 \times \mathbb{R}$, see in Fig.~\ref{fig02}.  One gets the same homotopy as for $\mathcal{M}_t$, namely: 
 \begin
{equation}
\label{e3}
  \pi_{i}(\mathcal{M}_{r})\sim \pi_{i}(S_{2}) \rightarrow \\ 
   \pi_{1}(\mathcal{M}_{r})\sim \{1\},\,\pi_{2}(\mathcal{M}_{r})\sim \mathbb{Z},\,\pi_{3}(\mathcal{M}_{r})\sim \mathbb{Z},\,\pi_{4}(\mathcal{M}_{r})\sim \mathbb{Z}_2.
\end
{equation}
Hence there are both Volterra continuous and topologically stable defects, whose dimensionalities in $M^4$ are: 1 for $\pi_{2}(\mathcal{M}_{r})$ and 0 for $\pi_{3}(\mathcal{M}_{r})$. $\pi_{4}(\mathcal{M}_{r}) $ designates possible \textsf{configurations}.
All line defects are continuous defects. By definition, they obtain by a VP performed on the full spacetime, defined by any element of the subgroup $P_{r}\sim H_r\ltimes \mathbb{R}^{1,3}$. Notice that these lines defects appear here as the continuous defects of a 2D euclidean plane or any homotopic manifold, but there is no such manifold of easy physical significance in the spacetime; it is therefore more sensible to attach the $H_r$ Volterra forms to the full group of isometry of the spacetime.
\\ \vspace{-10pt}
  
\textsf{4- Vanishing momenta} 

$p^a=0$: the little group is the full proper Lorentz group $L_0$; the coset space $\mathcal{M}_{v} = L_0/L_0$ is reduced to a point. Therefore all the homotopy groups $\pi_n(\mathcal{M}_{v})$ are trivial and the only possible defects are continuous VP defects. All the elements of $L_0$ have been scanned with the previous little groups, and thus this case embraces all the forms already cited. 
\vspace{-10pt}

 \subsection{The significance of topological defects: $M^4$-crystals, QFT vacuum; cosmological principle} \label{stab-cryst} \vspace{-10pt}
  
  We believe that it does not make sense of considering defects of a pure space or spacetime which does not contain any substance (energy or/and matter). The foregoing classification suggests two possibilities: 
  
1$-$ \underline{The first three conjugacy classes}.  In the spirit of \textcite{wigner39}, the relationship between elementary particles and conjugacy classes of $L_0$ subgroups suggests to identify a set of identical elementary particles $-$ of the same 4-momenta ${p}^a$ in $M^4$, of constant density, represented by some subgroup $H$ of $L_0$ leaving ${p}^a$ invariant, forming a \textit{congruence of geodesics} along ${p}^a$, $-$ to a \textit{crystal in $M^4$}, whose symmetry group is precisely the little group that leaves $p^a$ invariant. According to the discussion above, the first three conjugacy classes yield three types of crystals, with inhomogeneous symmetry groups $P_m$, $P_t$, and $P_r$ \footnote{We leave aside the question of the physical reality of tachyons.}. Thereby the topological and Volterra defects of these crystals would be those defined in the previous section. The use of 4-momenta to getting the three types of continuous forms is therefore fully justified.  The possible physical existence of $M^4$-crystals of massive or massless particles is discussed to some extent in \cite{kleman09}. 
  
   2$-$ \underline{The fourth conjugacy class} appears here in a new perspective, not so trivial. 
The $m$-, $r$- and $t$-forms it embraces could be observed if there is some substance that possesses all the $L_0$ symmetries, a sort of spacetime amorphous substance. We identify tentatively this substance with a QFT vacuum.\\   \vspace{-10pt}
 
It is worth analyzing this classification with regard to the cosmological principle.
In its usual acception, this principle states that any two elements belonging to the same spacelike slice are strictly equivalent (at least at some large enough scale); matter and energy, as a whole, are homogeneous and isotropic in 3D.
This \textit{narrow} cosmological principle (nCP) is in contrast with
a \textit{perfect} cosmological principle (pCP) $-$ utilized in the spacetime models of Bondi, Gold, and Hoyle \cite{bondi48,*hoyle48} $-$ which states that any two events in 4D are strictly equivalent.
Expressed in terms of symmetry groups, a)- nCP states that the spacetime is foliated by  maximally symmetric spacelike hypersurfaces parameterized by the cosmic time \cite{weinberg72}, whose Volterra defects are related to its group of isometries (e.g., $nCP \, \sim SO(3)\ltimes\mathbb{R}^{3}$ in $M^4$); b)- pCP implies a physical situation invariant under the Poincar\'e group $pCP \sim P(4)\equiv L_0 \ltimes \mathbb{R}^{1,3}$ (in $M^4$) or the de Sitter group $pCP \sim SO(1,4)$ (in $dS_4$). 
  
  Clearly, the only nCP-compatible forms are the $m$-forms, belonging to $m$-crystals. This does not preclude the existence of $r$- and $t$-forms, probably under the status of imperfect defects, see Sect.~\ref{obscut}. On the other hand all three types of forms are pCP-compatible, which stresses the interest of the QFT vacuum hypothesis above. Because the vacuum relevant forms are attached to pCP, they are not valid physical defects at the present epoch; but pCP defects may be associated with a de Sitter universe, whose existence is advanced in the inflationary theory of the primeval Universe, see Sect.~\ref{forms-net}.\vspace{-10pt}
  

\subsection{Forms in $M^4$ $vs$ defects in $E^3$}\label{boundcond} \vspace{-10pt}
 It is generally agreed that a GR singularity is identified by an \textit{incomplete geodesic}, meaning that such a geodesic runs into a \textit{hole} of the spacetime in a finite amount of time (or a finite variation of the affine parameter for a null geodesic),
 or took birth in a hole a finite time ago, $-$ in which hole the spacetime is singular, i.e. not properly defined $-$, see \cite{wald84}, chapter 9.  No doubt that \textsl{topological} $r$-forms, as far as they exist, are accompanied by such singularities; the question arises whether it is the same for \textsl{continuous} Volterra defects. 

\begin{figure}[h]
\includegraphics[width=1. in]{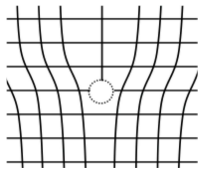}
   \caption{ 2D cut orthogonal to an edge dislocation line L in a square lattice. The line intersections represent atoms. Observe the presence of incomplete rows of atoms hitting the core region (the interior of the dotted circle) where the 'order parameter' of the crystal is no longer defined.}
 \label{fig03} 
\end{figure}

The foregoing definition of a spacetime singularity is perfectly compatible with the image we have of a \textsl{topological} Volterra defect in CMP, i.e., when the defect belongs to a non-trivial class of $\pi_1(\cal M)$; the VP can be smoothly achieved on the cut surface $\Sigma$, but has to stop short at a small distance of the line L$=\partial \Sigma$ itself, where it is not defined. It thus appears incomplete rows of atoms $-$ the equivalent of incomplete geodesics $-$, and the usual elasticity equations are not valid in a small \textsf{core} region about the line $-$ the equivalent of the hole. The core region has the topology of a toric cavity L$\times S^1$. Fig.~\ref{fig03} represents a 2D cut of a straight \textsl{edge} dislocation which displays such an incomplete row of atoms. 

Consider now a \textsl{continuous} Volterra defect which belongs to the trivial class of $\pi_1(\cal M)$; the Burgers circuit would not measure a nil displacement vector, one still expects incomplete rows of atoms that would emphasize the edge character of dislocations, or the twist character of disclinations, but the core can vanish away if this displacement vector dissipates into Volterra defects with infinitesimally small displacements, which is topologically allowed and physically feasible by an irreversible viscous dissipation process in CMP. At low temperatures this process might be little activated, and a singularity still be visible. Likewise, one expects incomplete geodesics in GR continuous defects, but it is not clear whether a comparable spread of the core exists. Actually, the gravitational forces may act in an opposite direction and (meta)stabilize a core singularity for a continuous defect, on account of an entropy increase! \vspace{-10pt}
    
  \section{Continuous cosmic forms} \label{forms} \vspace{-10pt}

We center our discussion on \textit{disclinations}, their associated metrics and structural properties. Other aspects (energetic and dynamical properties, detailed structure of the core) are left outside the present considerations. \textit{Dislocation} structural properties, whose necessity is not assessed in the frame of the field theory of particle transitions, should vary much from one type of spacetime to another. Contrarywise disclinations are related to the $L_0$ subgroup, which is present in all spacetimes. Therefore the following results, which are for disclinations in $M^4$, are valid for any spacetime.

However dislocations are present even in this restricted context. In particular they play a distinctive role in all processes that involve movement, change of shape, relaxation, etc. of disclinations, somewhat analogously to the role they play in CMP \citep{klfr08}. Also they carry torsion but not curvature. 

The main concepts related to cosmic forms will be developed for the $m$-form (the classical cosmic string of the present literature), and will not be detailed again for the $t$- and the $r$-forms. However the metric of the $r$-form, which has never been considered, to the best of our knowledge, will be given a fuller account.

Having in view the crystal interpretation of the $m$-, $r$- and $t$-forms, we also comment on the \textit{observability} of the cut hypersurfaces $\Sigma_m$, $\Sigma_t$, or $\Sigma_r$ about which the VP is performed: insofar as the spacetime distribution of 4-momenta displays the same symmetry as that one carried by the cosmic form, \textit{the cut hypersurface is not observable}. This is not so if the symmetry of the cosmic form is not respected by the 4-momentum distribution.

We first recall the $S\ell(2,\Cc)$ representation of the Lorentz group, also used in  \cite{wigner39}. \vspace{-10pt}

\subsection{The $S\ell(2,\mathbb{C})$ representation of the Lorentz group $L_0$, a reminder}\label{PGr} \vspace{-10pt}

$L_0$ as a manifold is connected but not simply connected, (its fundamental group is $\pi_1(L_0)\sim \mathbb{Z}_2$); denote $\widetilde{L_0}$ its simply connected double cover, $\pi_1(\widetilde{L_0})\sim \{1\}$. $\widetilde{L_0}$ is isomorphic to the special linear group $S\ell (2,\mathbb{C})$, which is the group of 2$\times$2 matrices 
$A=\begin{array}{|cc|}
  a    &   b \\
   c   &   d \\
\end{array}$
with complex entries and unit determinant $ad-bc=1$.  

A Lorentz transformation acting on an event
 $X^a=\{X^0,X^1,X^2,X^3\}$ $\in M^4$ is written as:\begin{equation}
\label{e4}
X\mapsto x=AXA^*,
\end{equation} where \begin{equation}
\label{e5}
{{X}}
=\begin{array}{|cc|}
 X^0+X^3     &  X^1-i\,X^2  \\
  X^1+i\,X^2    &   X^0-X^3
\end{array}\end{equation} 
is a 2$\times$2 hermitian matrix
and $A \in Sl (2,\mathbb{C})$; $A^*=\overline{A^t}$ is the conjugate transpose of $A$. 
$A$ and $-A$ yield the same result, which heals the problem raised by the 2:1 homomorphism (the \textit{spinor map}) introduced above \begin{equation}
\label{e6}
\mathrm{Spin}:\,S\ell(2,\mathbb{C})\rightarrow {L_0}.
\end{equation} 
Notice that we have \begin{equation}
\label{e7}
\mathrm{det}(X)\equiv (X^0) ^2 -(X^1) ^2-(X^2) ^2-(X^3) ^2=-\|{X^a}\|^2
=-\|{x}^{a}\|^2=\mathrm{det}(x),\end{equation}
with 
 $\|{x}^a\|^2=x^\alpha\,x_\alpha $ is $<0$ for ${x}^a$ timelike, $>0$ for ${x}^a$ spacelike, $=0$ for ${x}^a$ null.

Eq.~\ref{e6} maps the unitary subgroup $SU(2)\subset S\ell(2,\mathbb{C})$ onto the rotation subgroup $SO(3) \subset L_0$. 
 An element $A\in SU(2) \sim \widetilde{SO(3)}$ can be written in matrix form $ \begin{array}{|cc|} a     &  b  \\  -\bar{b}    &   \bar{a} \end{array},\, \|a\|^2+\|b\|^2=1$. \vspace{-20pt}

 \subsection{$m$-cosmic forms}\label{mforms} \vspace{-10pt}
    
      A static $m$-cosmic form with cylindrical symmetry about a timelike 2D plane $\Omega$, the \textit{world sheet}, 
      is the analog of a wedge disclination, i.e. a defect with cylindrical symmetry in a 3D space. 
      $\Omega$, which is the singular part of the $m$-form in the spacetime, will be taken along the $\{ \textbf e_0 \,\textbf e_3 \}$ plane. We shall not consider the (mathematically feasible) situation where the singularity is spacelike, in which case the form would be an hypersurface in a 3D space, not a line, cf. \cite{puntigam97}. \\ \vspace{-10pt}
      
 \textsf{1- The wedge $m$-form}

 Let $p_m^a $ be some timelike 4-momentum; it is invariant under any rotation about itself, and about any rotation $g \in H_m \sim SO(3)$ that leaves invariant the spacelike manifold ${P}_{m\bot}$ orthogonal to $p_m^a $. $H_m$ is the little group alluded to above, Sect.~\ref{cosmicforms}. Thus $g$ defines a global rotation of ${P}_{m\bot}$  about some spacelike rotation axis $\bm \varpi  \subset {P}_{m\bot}$. The timelike plane $\Omega \equiv \{ \textbf p_m^a  \,\bm \varpi \}$ is thus a 2-plane of rotation.
 
To make things simple, take $p_m^a = \{p_m^0,0,0,0\}$; thus ${P}_{m\bot} \equiv \{ \textbf e_1 \,\textbf e_2 \,\textbf e_3 \} $. Any element of $\widetilde{H_m}$ has the form $A_m=\scriptsize{\begin{array}{|cc|} a & b\\  -\bar{b} & \bar{a} \end{array}} \in SU(2) \sim \widetilde{SO(3)}$ \cite{wigner39,sternberg94}, 
 when  $p_m^a$ is as defined. The displacement of a point
$X_{m}^a=\{X_m^0,X_m^1,X_m^2,X_m^3 \}$ on the cut hypersurface $\Sigma_m$ 
reads $d_m =x_{m}-X_m = A_m \, X_{m} \,A_m^*-X_m$. 
$A_m$ rotates altogether any point $\textbf{X}_m \in \Sigma_m ,\, \textbf X_m \rightarrow \textbf{x}_m$ and the  causal structure attached to it about the 4-momentum ${p}^a_m$. We eventually get, with 
$ a=\cos \psi\, e^{-i\frac{\alpha}{2}},\, b=\sin \psi\,e^{-i\frac{\beta}{2}}, \, \rho e^ {i\theta} = x_m^1+i\,x_m^2,  \,u= \frac{\alpha -\beta}{2}+\theta,\, $ $\omega =\frac{\alpha +\beta}{2}:$  
$${x}_m^{1}+i{x}_m^{2}=e^{i \omega}\left[\rho\, (\cos 2\psi\, \cos u +i \sin u)-X_m^3\,\sin 2 \psi \right],$$
\begin{equation}
\label{e8}
{x}_m^0=X_m^0,\quad {x}_m^{3}=\rho\,\sin 2\psi\,\cos u + \,X_m^3 \cos 2 \psi;
\end{equation} $\bm \varpi$ can be obtained from this equation.

This is the most general case for $p_m^a = \{p_m^0,0,0,0\}$. 
For the sake of simplicity we assume $a=\exp- i\alpha/2,\,b=0$ (which yields $\{\bm \varpi \} =\{ \textbf e_3\}$);
  $\Omega \equiv \{\textbf e_0\,\textbf e_3\}$, the plane of rotation, is invariant under the action of $A_m$; the $\{\textbf e_1\,\textbf e_2\}$ plane is translationally invariant, and: 
  \begin{equation}
\label{e9}
{x}_m^0=X_m^0,\quad {x}_m^3 = X_m^3,\quad {x}_m^1+i\,{x}_m^2 =({X}_m^1+i\,X_m^2)\exp{i\alpha}. 
\end{equation} 
 The $ \{\textbf e_3 \}$ axis is an axis of rotation in the spacelike submanifold $P_{m\bot} $ and the $\{\textbf e_0 \}$ axis in the timelike submanifold $\{ \textbf e_0\,\textbf e_1\,\textbf e_2 \}$. We haven't yet selected a particular world sheet; take it now along the 2-plane axis of rotation $\Omega$; the restriction of the singular part of the $m$-form to the spacelike manifold $\{ \textbf e_1\,\textbf e_2\,\textbf e_3 \}$ is therefore this $\{ \textbf e_3 \}$ axis. By analogy with the CMP disclination, this will be called a \textit{wedge $m$-form}. 
  
  This {wedge} $m$-disclination is the same as the conical cosmic string of ref. \cite{vilenkin81b,*hiscock85,*tod94}; Tod also defines the conical cosmic string as a straight disclination line along the $\{\textbf{e}_3\}$ axis at constant cosmic time. 
 In fact, all the displacements take place in a 3-space orthogonal to $p_m^a $, and result from the action of the $SO(3)$ selected group element. Therefore the analysis of the cosmic strings in space does not differ from the analysis of continuous disclinations in a Euclidean 3-space inhabited by an homogeneous and isotropic substance, \textit{i.e.}, from the analysis of the disclinations in a glass. And since wedge disclinations in space can take any shape of a closed loop (or go to infinity), similarly $m$-disclination loops can take any shape (or go to infinity).  \\ \vspace{-10pt}

 \textsf{2- The metric of a wedge $m$-form}\label{m-metric}  

The metric of a straight disclination with rotational symmetry about the $\{\textbf{e}_0 \,\textbf{e}_3\}$ world sheet obtains directly by applying the Volterra process to the spacetime, \textcite{tod94}.  This requires a simple change of coordinates of the line element of a pure Minkowskian spacetime 
\begin{equation}
\label{e10}
dS^2 = -dT ^2+dR^2 + R^2 \,d\Phi^2 +dZ^2, 
\end{equation}namely 
\begin{equation}
\label{e11}
t=T, \quad r = R, \quad \phi =  \frac{\Phi}{1 - \frac{\alpha}{2\, \pi} },\quad z = Z,
\end{equation}
 which expresses that the angle $\Phi$ is dilated (for $\alpha > 0$) or compressed (for $\alpha < 0$) by the factor $a^{-1}, \, a=1 - \frac{\alpha}{2\, \pi}$, when running from $\Phi = 0$ to $\Phi = 2\, \pi - \alpha$.  One gets:
\begin{equation}
\label{e12}
ds^2 = -dt^2 +dz^2 + dr^2 + a^2\,r^2 \,d\phi^2,
\end{equation}
 $\alpha$ being the angle occurring in the Volterra process. The Riemann tensor vanishes everywhere in the spacetime, except on the world sheet or some hypersurface $\Sigma$ separating the 'core' region from the bulk of the cosmic string, see below. 

The change of coordinates (Equation \ref{e11}) used to obtain Equation \ref{e12} for a wedge $m$-form, viz the dilation of the angle $\Phi$, can also be written, in a more easily generalizable shape: 
\begin{center}${x}_m^0=X_m^0,\quad {x}_m^3 = X_m^3,\quad {x}_m^1+i\,{x}_m^2 =({X}_m^1+i\,X_m^2)\exp{i \frac{\alpha\, \phi}{2\, \pi}}, \qquad$i.e.\end{center} \vspace{-10pt}
\begin{equation}
\label{e13}
{x}^1= X^1 \,\cos  \frac{\alpha\, \phi}{2\, \pi}  - {X}^2 \,\sin   \frac{\alpha\, \phi}{2\, \pi} , \quad {x}^2= X^1 \,\sin  \frac{\alpha\, \phi}{2\, \pi}+ {X}^2 \, \cos   \frac{\alpha\, \phi}{2\, \pi}.
\end{equation}

Equation \ref{e13} obtains by interpreting Eq.~\ref{e9} as giving the displacement on $\Sigma_m$ after the traversal of a loop circling about the 2-plane of rotation $\{ \textbf e_0\,\textbf e_3 \}$ when $0 < \phi \leqslant 2\, \pi$. 

Notice that Eq.~\ref{e12} can also be written:
\begin{equation}
\label{e14}
ds^2 = -dt^2+(dx+\frac{\alpha}{2\,\pi}\,y \,d\phi)^2 + (dy-\frac{\alpha}{2\,\pi}\,x \,d\phi)^2 +dz^2,
\end{equation} where the rotation about the $\{\textbf e_3\}$ axis is apparent.\\\vspace{-10pt}

According to the sign of $\alpha$, there are two types of disclinations:

$-$ {$\alpha >0, \, a < 1$}; a sector of angle $2\, \pi \,(1-a)$ is cut out from the spacetime and the resulting edges are identified; $\alpha$ is a \textit{deficit angle}. In the terminology of CMP, this is a \textit{positive disclination}. All cosmic strings are of this type,

$-$ {$\alpha < 0, \, a > 1$}; a sector of angle $2\, \pi \,(a-1)$ is inserted in the spacetime and its edges are identified with the edges of the cut that has been practiced. In the terminology of CMP, this is a \textit{negative disclination}.\\ \vspace{-10pt}

Curvature is associated with the stress-energy tensor $T_{\alpha \beta}$ through the Einstein's equation. Thus the mass-energy $\mu$ per unit cosmic line length is related to the curvature of a $m$-form \cite{vilenkin81b,gott85,hiscock85,tod94}, \cite{langer70,*linet85}. We recall that 
 \begin{equation}
\label{e15}
\mu = \frac{c^2}{4\,G}\,({1-a})\equiv   \frac{c^2}{8\,\pi \,G} \,\alpha,
\end{equation} where $G$ is the gravitation constant and $c$ 
the light velocity \footnote{
With $ {c^2}/{8\,\pi \,G}\approx 0.51 \times 10^{26}$ kg/m and $  \alpha \approx 10^{-6}$ rad, a GUT cosmic string has a mass of $\approx 10^{20}$ kg/m. A GUT string long enough to cross the observable Universe would weight about $10^{16}\, M_{\odot} $, which is the typical mass of a galaxy.}.  

Two remarks are in order:

$-$ The positivity of the mass-energy forbids the existence of negative disclinations. However this is without taking into account the possibility of multiple vacua of positive as well as negative energies. See the discussion in Sect.~\ref{disc}.

$-$  Eq.~\ref{e15} can be derived for a $m$-form limited to the outer part of a timelike circular cylinder whose axis is the world sheet, the inner part (the 'core') being a Minkowskian spacetime without any type of singularity. The result is independent of the radius of the cylinder \cite{langer70}.\\ \vspace{-10pt}

 \textsf{3- The twist $m$-form}

Consider now a disclination (denoted in \cite{puntigam97} as \textsl{"disclination 5"}). It is a $m$-form whose 2-axis of rotation is now $\{\textbf e_0 \, \textbf e_2\}$. It has the same world sheet $\{\textbf e_0 \, \textbf e_3\}$ as above. This world sheet is not invariant, even globally, under the action of the VP. This has some important consequences, as we see below. In the 3 space $\{\textbf e_1 \, \textbf e_2 \,\textbf e_3\}$ the cosmic form is along the $\{\textbf e_3\}$ axis, and the rotation about the $\{\textbf e_2\}$ axis, orthogonal to it. In CMP a similar situation would be called a \textsf{twist disclination}; we keep the same terminology.

The variation of the rotation when circulating around the world sheet can be written:
\begin{equation}
\label{e16}
 x = X^1 \,\cos \frac{\alpha}{2\, \pi}  \, \phi - X^3 \,\sin\frac{\alpha}{2\, \pi}  \, \phi  , \quad z= X^1 \,\sin \frac{\alpha}{2\, \pi}  \, \phi + X^3 \, \cos \frac{\alpha}{2\, \pi}  \, \phi ,  \quad r=R \quad \phi = \Phi, \end{equation} 
and the line element takes the form (with more usual notations):
\begin{equation}
\label{e17}
ds^2 = -dt^2 +(dx+\frac{\alpha}{2\, \pi} \,z \,d\phi)^2 + dy^2 +(dz - \frac{\alpha}{2\, \pi} \,x \,d\phi)^2,
\end{equation} \noindent where $d\phi$ is as above in Eq.~\ref{e14}.\\ \vspace{-10pt}

\textsl{The singular world shell of a twist $m-$form}

This cosmic form exhibits a property not yet met, and which has no equivalent in CMP; 
the singularity is not carried by the world sheet $\{\textbf e_0 \, \textbf e_3\}$ but by a timelike world shell $L_m \equiv \{\Hcal_m=0\},$ \begin{equation}
\label{18}
\Hcal_m=r-\frac{\alpha}{2\, \pi}\,z\,\sin\phi.
\end{equation} The non-vanishing Christoffel symbols, inverse metric components, all behave as  $\Hcal_m^{-1}$ and thus diverge on $\Hcal_m=0.$ The geodesics get singular on the world shell \footnote{This calculation, as others in this paper (Christoffel symbols, Riemann tensor, geodesics $\cdots$) is done with the help of the \textsc{mathematica} notebook of App. C in \cite{hartle}. This author will gladly provide detailed results on demand.}.
  
Therefore there are incomplete geodesics that terminate (or start) on the world shell, (in the wedge case geodesics terminate (or start) on the world sheet). The metric of Eq.~\ref{e17} being valid in the inner part of the world shell as well as the outer part, there are such geodesics on both sides of the hypersurface $L_m$. 
\\ \vspace{-10pt}

The appearance of a singular world shell has some mathematical advantages. \textcite{geroch87} have argued that postulating a Dirac distribution of curvature on a 2D manifold is not a well-posed mathematical problem, but that there is no such trouble for a 3D one. In fact, whereas the inverse metric $g^{\alpha \beta}, \,\alpha, \beta \cdots = 0,1,2,3$ is singular on $L_m$, this is not so for the metric $g_{ij}$ and the inverse metric $g^{ij},\, i,j \cdots = 1,2,3$ restricted to $L_m$, as well as its extrinsic curvature \cite{barrabes91}. The metric can be written
$$g_{ij} =e_i^\alpha\,e_j^\beta \, g_{\alpha \beta},$$ where the $e_i^\alpha$ form a coordinate basis on $L_m$, such that $e_i^\alpha =\frac{\partial x^\alpha}{\partial \xi ^i};$ the $x^\alpha$ are general coordinates, and the $ \xi ^i$ are coordinates on $L_m$.

Adopting cylindrical coordinates $x^\alpha =\{t,r,\phi,z\}$ (in this order) for the $x^\alpha$, and the coordinate basis:
$$e_1^a =\{1,0,0,0\},\quad e_2^a =\{0,\frac{\alpha}{2 \pi}\,z \,\cos \phi,1,0\},\quad e_3^a =\{0,\frac{\alpha}{2 \pi}\,\sin \phi,0,1\},$$ on $L_m$ (these three independent vectors are tangent to $L_m$ and derive from a set of coordinates $\xi^i =\{t,\phi,z\}$ $-$ in this order), then $g_{ij}$ and $g^{ij}$ are regular metric tensors. We do not explore the properties of the hypersurface $L_m$ in the present article.\\ \vspace{-10pt}
 
 \textsf{4- Relation between twist cosmic $m$-forms and dislocations}
   
  The world sheet and the world shell $L_m$ are both not translationally invariant along the $\{\evec_3\}$ axis of the cosmic form. A somewhat similar situation occurs with CMP twist disclinations, as it is visible on Fig.~\ref{fig01}d: the singular line (the border of the cut surface) loses translational invariance when it experiences the VP. Of course this situation is not entirely satisfactory and certainly highly energetic. It has been analyzed (and cured) in CMP. A comparable analysis works for the twist $m$-form.  
 
  $-$ \underline{Twist disclination lines in CMP}: \begin{figure}[h]
\includegraphics[width=2.8 in]{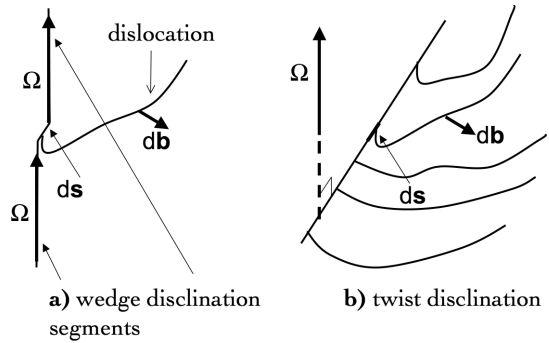}
   \caption{a)- Kink $d\svec$ along a CMP wedge disclination of strength $\bm \Omega$; b)- twist disclination made of an assembly of such kinks; a few attached infinitesimal dislocations, with the same Burgers vector density $d\bvec / d\svec$, are indicated. The whole set of these dislocations spreads through space, generically; here we have sketched them as distributed on a surface bordered by the twist line, which does not implies any lack of generality for the interactive topological properties between the main disclination and the attached densities.}
 \label{fig4} 
\end{figure} it can be shown \cite{klfr08} that infinitesimal dislocation densities can attach to a disclination line, to the effect that the line gets curved and in fact takes any shape away from the wedge geometry (which does not require attached dislocations) for a suitable dislocation density. Let $\alpha$ be the (small) deficit angle characterizing the disclination, $\bm \Omega$ along the axis of rotation with $|\bm \Omega| = \alpha$, then a kink of the disclination line along an infinitesimal segment $d\svec$  Fig.~\ref{fig4}a can be interpreted as resulting from the attachment at the location of $d\svec$ of a dislocation having an infinitesimal Burgers vector 
  \begin{equation}
\label{e19}
d\textbf b = \bm \Omega \times d\svec.
\end{equation} Notice that the axis of rotation $\bm \Omega$ is shifted by the vector $d\svec$ and thus keeps aligned with the wedge segments. 
For a detailed discussion of Eq.~\ref{e19} and its physical implications, see \cite{klfr08}.

Any curved disclination obtains by adding such infinitesimal dislocations along the line, with suitable densities. Observe that, generically, one expects that such densities are spread through all space. Fig.~\ref{fig4}b illustrates the case of a \textit{constant} dislocation density, and for  the sake of simplicity it is assumed that the densities are distributed on a surface, a situation which is not very likely if the attached defects carry infinitesimal strengths. But the distribution of the densities depends on the boundary conditions, while their topological content is an invariant expressed by the Eq.~\ref{e19}. One gets a twist disclination if $d\svec \cdot \bm \Omega = 0$.

There is ample experimental evidence  in liquid crystals of the soundness of this analysis, first incentivized by the observation of curved and closed disinclination loops in cholesterics \cite{friedel69}, thereby in contrast with the naive prediction of wedge only disclinations. Also, some optical microscopy observations in liquid crystal columnar phases have visualized a situation much akin to that one of Fig.~\ref{fig4}a, with two half wedge disclinations shifted by a kink carrying a set of dislocations (Fig. 3 of \cite{kln}).

$-$ \underline{Cosmic forms in GR}: 
Consider first an edge dislocation with world sheet $\{\textbf e_0 \, \textbf e_3\}$ as above, the Burgers vector $\textbf b$ along the $\{ \textbf e_1\}$ direction. Its line element, in Cartesian coordinates $\{t,x,y,z\}$, is:
\begin{equation}
\label{e20}
ds^2 = -dt^2+(dx+\frac{b_x}{2\,\pi}\,d\phi)^2 +dy^2 +dz^2. 
\end{equation}  Here again it appears a hypersurface $L_e \equiv \{\Hcal_{e_x} = 0\}, \, \Hcal_{e_x} =r-\frac{b_x }{2\,\pi} \, \sin \phi$ along which 
some non-vanishing Christoffel symbols, inverse metric components, and  terms appearing in the geodesic equations behave as  $\Hcal_{e_x}^{-1}$ and thus diverge on $L_e$. The presence of a world shell (now translationally invariant) in this edge dislocation, noticed incidentally by \textcite{krasnikov07}, has no equivalent in CMP. In CMP, the inner core region along the line would relax elastically to a cylinder of radius approximately equal to $b_x$ (as here; the radius of $L_e$ is $|b_x|/(4 \, \pi)$) but with cylindrical symmetry about the $\{\textbf e_3\}$ axis (at least in the case of isotropic elasticity), and the core region would exhibit a different 'order parameter' than the outer region.

But these differences being made, the same topological property that relates a dislocation to a disclination in CMP is present in GR. Compare indeed $\Hcal_m=r-\frac{\alpha }{2\,\pi}\,z \,\sin \phi$ and $\Hcal_{e_x}=r-\frac{b_x }{2\,\pi} \, \sin \phi$; these two expressions can be given the following interpretation: it is possible to attach a dislocation of Burgers vector $b_x$ to the cosmic form per length $\delta z= b/\alpha$. In other words (and passing to a continuous Burgers vector) the $z$ dependence of the world shell $L_{m}$ can be removed by the addition of a set of continuous dislocations, attached to the world shell of the cosmic form, with density $\alpha= db_x/dz$ per unit length of cosmic form. If one observes further that a dislocation can take any shape once its Burgers vector is fixed (as illustrated Fig.~\ref{fig4} for CMP), one is led to the following picture: a world shell $L_m$ being given, trace on it a set of timelike parallel lines $\ell_e$ at equal distances $ds$, then attach to each of those lines $\ell_e$ the world shell $L_e$ of a dislocation ($L_e$ leaning on $\ell_e$, but otherwise $L_e$ is not compelled to coincide with $L_m$), of Burgers vector $db$, then $L_m$ is transformed to a translationally invariant world shell. This process clearly stabilizes the twist cosmic form. The relationship $\alpha= db/ds$ is similar to Eq.~\ref{e19}. 

Notice that as a result of the attachment process the Riemannian curvature of the original spacetime is modified: the metric around the twist $m$-form is no longer Minkowskian. This is easy to understand: it is possible to choose without restriction, for the cut hypersurface of the twist line equipped with the attached defects, a manifold generated by these defects themselves. Then the total VP, including these attached defects, does no longer displace 'rigidly' the two lips apart. In fact the relative displacement of the lips varies from place to place (suffering a variation $d\bvec$ from one side of a dislocation to the other) and is no longer an element of symmetry. 

We do not investigate further in this article the detailed configuration of the attachment and the energy of the core, a topic in itself.
\vspace{-10pt}

\subsection{$t$-cosmic forms} \label{hypdiscli}  \vspace{-10pt}

 \textsf{1- The wedge $t$-form} 

   A \textit{Lorentz boost} is an imaginary rotation; with an hyperbolic angle $\omega$ we have: 
   \begin{equation}
\label{e21}  x = A_t\, X\, A_t^*
,  \qquad \mathrm{with}\qquad
A_t=\begin{array}{|cc|}
 e^{\frac{\omega}{2}}     &  0  \\
 0    &    e^{-\frac{\omega}{2}} 
\end{array} =A_t^*.  \end{equation}
We want to construct a wedge form with a VP characterized by $A_t$. No $t$-crystal with such a cosmic form is a priori liable to exist (no tachyons); however $t$-forms are possible forms of a spacetime amorphous substance, as mentioned in Sect.~\ref{stab-cryst}. The spacelike directions invariant under the action of $A_t$ can be written $p^a_t = \{0,p^1,p^2,0\}$, $\forall \, p^1,\, \forall \,p^2$.

 It is thus possible in a quadruple infinity of manners to choose a pair of independent spacelike directions $p^a_t$; such a pair would define a locally invariant 2-plane which could be taken as a world sheet for a wedge $t$-form carrying the hyperbolic rotation $A_t$. But this world sheet, being spacelike, is not an appropriate line defect in space; however there are globally invariant 2-planes, like 
$\{\textbf e_0 \, \textbf e_3\}$ which are. \\ \vspace{-10pt}

\textsl{The metric of a wedge $t$-form}

An hyperbolic rotation in the $\{\textbf e_0\,\textbf e_3\}$ plane reads
 \begin{equation}
\label{e22}
{x}^0= X^0 \,\cosh \omega+{X}^3 \,\sinh \omega , \quad {x}^3= X^0 \,\sinh \omega+ {X}^3 \, \cosh \omega , \quad {x}^1=X^1, \quad {x}^2=X^2, \end{equation} and indeed leaves this plane globally invariant.

The same method as above that yields Eq.~\ref{e13} extends to getting the line element of a wedge $t$-form, carried out as for the $m$-form on a cut hypersurface $\Sigma_t$ bordered by the world sheet $r = 0$ (the plane $\{\textbf e_0 \, \textbf e_3\}$). One can write (with new notations  $\{t,r,\phi,z\}$ adapted to cylindrical coordinates):
\begin{equation}
\label{e23}
 t = T \,\cosh \varpi  \, \Phi + Z \,\sinh \varpi  \, \Phi  , \quad z= T \,\sinh \varpi  \, \Phi+ Z \, \cosh \varpi  \, \Phi ,  \quad r=R \quad \phi = \Phi \end{equation} where $ \varpi = \frac{\omega}{2 \, \pi}$,
(the hyperbolic angle $\varpi  \, \phi$ varies between $0$ and $\omega$ when traversing a loop that surrounds the world sheet). Hence \cite{tod94,puntigam97}:
\begin{equation}
\label{e24}
ds^2 = -(dt - \varpi \,z \,d\phi)^2+dr^2 + r^2 \,d\phi^2 +(dz - \varpi \,t \,d\phi)^2.
\end{equation}

The Riemann tensor attached to the line element Eq.~\ref{e24}
vanishes identically, as expected from a Volterra process. \textcite{tod94} has shown that this disclination carries distributional curvature and torsion in the $\textbf{e}_0 \, \textbf{e}_3$ plane. 


 \textsf{2- The twist $t$-form; attached dislocations}\label{coret}

The disclination (denoted in \cite{puntigam97} as \textsl{"disclination 7"}) is a $t$-form with the same world sheet $\{\textbf e_0 \, \textbf e_3\}$ as above, and whose hyperbolic rotation (the Lorentz boost) is in the $\{\textbf e_0 \, \textbf e_2\}$ plane. Therefore the world sheet is not invariant in the VP. In the 3 space $\{\textbf e_1 \, \textbf e_2 \,\textbf e_3\}$ the cosmic form is along the $\{\textbf e_3\}$ axis, and the hyperbolic rotation along the $\{\textbf e_2\}$ axis, orthogonal to $\{\textbf e_3\}$. We keep the same terminology as above of \textsf{twist} disclination.

The variation of the boost when circulating around the $\{\textbf e_0 \, \textbf e_3\}$ plane can be written:
\begin{equation}
\label{e25}
 t = T \,\cosh \varpi  \, \Phi + Y \,\sinh \varpi  \, \Phi  , \quad y= T \,\sinh \varpi  \, \Phi+ Y \, \cosh \varpi  \, \Phi ,  \quad r=R \quad \phi = \Phi, \end{equation} 
and the corresponding line element is (notice a correction of sign with respect to \cite{puntigam97}):
\begin{equation}
\label{e26}
ds^2 = -(dt - \varpi \,y \,d\phi)^2+(dy- \varpi \,t \,d\phi)^2 + dx^2 +dz ^2.
\end{equation} 

The Riemann tensor vanishes identically. In analogy with the $m$-twist form, the inverse metric and the Christoffel symbols diverge on a hypersurface $L_t \equiv \{\Hcal_t =0\}$ with:\begin{equation}
\label{e27}
\Hcal_t = r - \varpi \,t\,\cos \phi.
\end{equation} 

Thus $L_t$ is not time-invariant. But, in analogy with the case of the twist $m$-form, this can be cured with the help of  a suitable dislocation density attached to the $t$-form as time passes by. The relevant dislocations are edge dislocations with the line element
\begin{equation}
\label{e28}
ds^2 = -dt^2+dx^2 +(dy+\frac{b_y}{2\,\pi}\,d\phi)^2 +dz^2. 
\end{equation}  The related singular hypersurface is $\L_{e_y} \equiv \{\Hcal_{e_y} =0\}$, with $\Hcal_{e_y} = r-\frac{b_y }{2\,\pi} \, \cos \phi$. And again in analogy with
the twist $m$-case, one can attach a dislocation density $db_y= \varpi \, dt$ to the twist form as it proceeds through time, yielding time invariance. \vspace{-20pt}

\subsection{$r$-cosmic forms}\label{forms-light}
  \vspace{-10pt}


The general expression for an element $A_r$ of $\widetilde{H_r}$, which leaves locally invariant the null direction $k^a =\{1,0,0,1\} \propto p_r^a =\{p,0,0,p\}$, can be written \citep{wigner39,sternberg94}: 
\begin{equation}
\label{e29}
A_r= \begin{array} {|cc|}
  \exp{-i\frac{\alpha}{2}}   \, \,\,&\,  -\bar{\sigma}\,\exp{i\frac{\alpha}{2}} 
  \\ 0 \,\,&\,   \exp{i\frac{\alpha}{2}} \end{array},\quad \mathrm{with} \quad\sigma = \sigma_1 + i\,\sigma_2.
\end{equation} No other null direction is invariant under the action of $A_r$. Therefore there is no world sheet locally invariant in any VP characterized by  $A_r$. On the other hand one can show that there is one 2-plane (and only one) that is VP globally invariant; this plane contains of course the null direction $k^a$. Choosing this plane as the world sheet would define a wedge $r$-form, in analogy with a wedge $t$-form, where the world sheet is only globally invariant.
  The null 3-manifold $P_{r\bot} \equiv x^0-x^3 = 0$ orthogonal to $k^a$ is globally invariant under the same action. This is in contrast with the $m$-case, where a 3D spacelike manifold is invariant; in that sense, a $r$-disclination does not obey the cosmological principle. We haven't carried out the calculation of the wedge $r$-form metric, which is rather cumbersome.\\ \vspace{-10pt}

Let $X^a =\{X^0,X^1,X^2,X^3\}$ be an event on the cut hypersurface (not defined yet) in $M^4$;
 its transform under $A_r$ is: 
 \footnotesize\begin{equation}
\label{e30}
\bm x = A_r\, \bm X \,A_r^* =\begin{array} {|cc|}
 X^0 + X^3 - X \, \bar \sigma \, \exp{i\, \alpha}- \bar X \, \sigma \, \exp{-i\, \alpha} +\sigma \,\bar \sigma \,(X^0 - X^3)&\quad\bar X \, \exp{-i\, \alpha} -\bar \sigma \,(X^0 - X^3) 
  \\  \quad X \, \exp{i\, \alpha} - \sigma \,(X^0 - X^3)   \,\,&\,   X^0 - X^3\end{array}.
\end{equation}
\normalsize 

This also reads: \footnotesize{
 \begin{equation}
\label{e31}
x^0 + x^3= X^0 + X^3 - \sigma \, \bar X \, \exp{-i \, \alpha}- \bar \sigma \, {X} \, \exp{i \, \alpha} +\sigma \, \bar \sigma \,(X^0 - X^3),\quad x^0 - x^3= X^0 - X^3,
\end{equation}
 \begin{equation}
\label{e32}
x =X \,\exp{i\, \alpha}-\sigma \,(X^0 - X^3),\quad \mathrm{where }\quad x =x^1 +i \,x^2,\,\, X =X^1 +i \,X^2.
\end{equation}} \normalsize
\indent Uppercase letters $X^i$ (resp. lowercase $x^i$) refer to the situation before (resp. after) the introduction of the disclination in $M^4$. We assume in the sequel that the world sheet of the $r$-disclination is the plane $\{\textbf e_0 \, \textbf e_3\}$ (not invariant uner $A_r$). A point in the plane
$\{\textbf e_1 \, \textbf e_2\}$ is defined by the coordinates $x^1, \,x^2$ or alternatively $r,\, \phi$. \\  \vspace{-10pt} 

 \textsf{1- The metric of a mixed $r$-form}\label{appli2}  

When traversing a loop which surrounds the world sheet, the angle $\phi$ scans the interval $\left[0 , 2\, \pi\right]$ whereas $\Phi$ scans the interval $\left[0, 2\, \pi - \alpha\right]$. The displacement on the cut hypersurface is the sum of a rotation by an angle $\alpha$ and of a translation $\sigma_1, \sigma_2$. Assuming proportionality in the increment of $\Phi$ when the loop is traversed, we have $\Delta \phi = \phi - \Phi = {\alpha \, \phi}/(2\, \pi)$. Eq.~\ref{e31} and \ref{e32} can be used not only on the cut hypersurface, but also all around the world sheet, with the following modifications: $\alpha$, the rotational component of the displacement is replaced by the increment $ \frac{\alpha \, \phi}{2\, \pi}$; the translational component is multiplied by  $ \frac{\phi}{2\, \pi}$, hence denoting
$$U=X^0- X^3, \,V = X^0+X^3,\,u=x^0- x^3, \,v = x^0+x^3,  \, x=x^1 +i\, x^2, \, w =  \frac{u \, \phi}{2\, \pi} =  \frac{U \, \phi}{2\, \pi},$$
the Eq.~\ref{e31} and \ref{e32} yield:
 \begin{equation} \label{e33}
v-V =- \sigma \, \bar{X} \, \exp{-i \, \frac{\alpha \, \phi}{2\, \pi}}- \bar \sigma \, {X} \, \exp{i \, \frac{\alpha \, \phi}{2\, \pi}} +\sigma \, \bar \sigma \,U\,\frac{ \phi}{2\, \pi}, \quad u-U = 0,\end{equation}
 \begin{equation}
\label{e34}
x = X \,\exp{i\, \frac{\alpha \, \phi}{2\, \pi}}-\sigma \,U\, \frac{ \phi}{2\, \pi},
\end{equation}
   These relations allow for the transformation of the line element $dS^2 = -dU\,dV + dX_1^2 +dX^2_2$ of the undeformed spacetime into the line element $ds^2 = -du\,dv + dx_1^2 +dx^2_2$ in the presence of the disclination. \\ \vspace{-10pt}

  i)- \textsl{The case $\alpha = 0, \, \sigma \neq 0 $}.  
  
The Eq.~\ref{e33} and \ref{e34} can be written:
\begin{equation} \label{e35}
v-V =- (\sigma \, \bar{X} + \bar \sigma \, {X} \, )\,\frac{\phi}{2\, \pi} +\sigma \,\bar \sigma \, w, \quad u-U = 0, \end{equation}
 \begin{equation}
\label{e36}
x = X -\sigma \,w, \end{equation}
Equation  \ref{e35} simplifies by taking for the cut hypersurface the timelike hypersurface $\Sigma_{r_{\alpha=0}}$ 
\begin{equation}
\label{e37}
\Sigma_{r_{\alpha=0}} \equiv \big\{ \half (\sigma \, \bar{X} + \bar \sigma \, {X} \, ) \equiv \sigma_1\, X^1 +\sigma_2\, X^2 = 0\big\}
\end{equation} bordered by the sheet $\{\textbf e_0 \, \textbf e_3\}$: according to the discussion in Sect.~\ref{obscut} below the result does not depend on the cut hypersurface, as long as the environment has the pCP symmetry $A_r$.

The line element $ds^2= -(dX^0)^2 +(dX^1)^2 +(dX^2)^2  +(dX^3)^2$, when expressed in function of the components $\{x^0,x^1,x^2,x^3\}$, contains the following terms: 
\begin{equation}
\label{e38}
 -(dX^0)^2 +(dX^3)^2 = -(dx^0)^2 +(dx^3)^2+ |\sigma |^2  \, \left[ (dx^0 - dx^3) \,dw \right].
\end{equation}
\begin{equation}
\label{e39}
 (dX^1)^2 +(dX^2)^2 = dX \,d\bar X = (dx^1)^2 +(dx^2)^2+ |\sigma |^2  \, dw^2+ 2 \, dw \, (\sigma_1 \,dx^1 + \sigma_2 \, dx^2).\end{equation} 
 Thus
 \begin{equation}
\label{e40}
ds^2 = -(d\lambda^0)^2 + (d\lambda^1)^2+(d\lambda^2)^2 +(d\lambda^3)^2,
\end{equation}
with \begin{equation}
\label{e41}
\lambda^0 =x^0 -\frac{|\sigma |^2}{2}\, w,\,\lambda^1 =x^1 +\sigma_1\, w,\,\lambda^2 =x^2 +\sigma_2\, w,\,\lambda^3 =x^3 -\frac{|\sigma |^2}{2}\, w.
\end{equation} 

 With these variables, the vanishing of the Riemann tensor is apparent. The search for a possible world sheet or world shell cannot be made with the coordinates $\lambda^a$, which are Minkowskian pure; we rather use the set of coordinates $u, \,v,\,r, \, \phi,$ 
 with which the line element can be written:
\begin{equation}
\label{e42}
 (ds)^2=  -du\,dv+ |\sigma |^2 \, du \,dw 
 + dr^2 + r^2 \, d\phi^2 + |\sigma|^2  \, dw^2+ 2 |\sigma|\, dw \, d\left[r\, \cos (\phi - \phi_0) \right], \end{equation} where we have used the notation
 $\sigma = |\sigma| \, \exp i \phi_0,$ i.e. $\sigma_1 \,x^1 + \sigma_2 \, x^2 = |\sigma | \, r \, \cos (\phi - \phi_0)$.

The Riemann tensor is vanishing everywhere, as expected, but might well be singular on the timelike 3-manifold $\Hcal_{r0} =0$:
 \begin{equation}
\label{e43}
\Hcal_{r0} =  r - \varsigma \,u\, \sin(\phi - \phi_0) , \qquad \mathrm{where}  \quad \varsigma = \frac{|\sigma|}{2\,\pi}.
\end{equation}
 $\Hcal_{r0}$ appears in the denominators of the inverse metric $g^{ab}$ $-$ except for $g^{u u} =0, \,g^{u v} =-2, \,g^{u r}=0,\,g^{u \phi} =0$ $-$, of the non-vanishing Christoffel symbols, and of the geodesics; the velocities become singular on $\Hcal_{r0} =0$.
  \\ \vspace{-10pt}

 ii)- \textsl{The general case $\alpha \neq 0, \, \sigma \neq 0 $}. 
 
The calculation uses the full Eq.~\ref{e33} and \ref{e34}; we assume that the cut hypersurface $\Sigma_{r_{\alpha}}$ is the hypersurface
\begin{equation}
\label{cch}
\Sigma_{r_{\alpha}} \equiv \big\{ \half (\sigma \, \bar{X}\, \exp{-i \, \frac{\alpha \, \phi}{2\, \pi}} + \bar \sigma \, {X} \, \exp{i \, \frac{\alpha \, \phi}{2\, \pi}} ) = 0\big\}.
\end{equation} The calculation of the line element requires some algebra but gives a rather simple result: 
\begin{equation}
\label{e45}
ds^2 = -(d\lambda^0)^2 + (\delta \mu^1)^2+(\delta \mu^2)^2 +(d\lambda^3)^2,
\end{equation} 
where  $\lambda^0$ and $\lambda^3$ are as above, Eq.~\ref{e41}, and \begin{equation}
\label{e46}
 \delta \mu^1 =d\lambda^1 + \frac{\alpha}{2\, \pi}\,\lambda^2\,d\phi,\qquad  \delta \mu^2 = d\lambda^2 -\frac{\alpha}{2\, \pi}\,\lambda^1\,d\phi,
\end{equation} where $ \delta \mu^1 $ and $ \delta \mu^2 $ are not total differentials. Eq.~\ref{e46} yields Eq.~\ref{e40} if $\alpha = 0$ and Eq.~\ref{e12} if $\sigma = 0$. The Riemann tensor is vanishing: the spacetime is Minkowskian, except where the Christoffel symbols and the differential equations for the geodesics are divergent. Using the same change of coordinates as above, one finds that the singularities are carried by the hypersurface $\Hcal_{r}=0$:
\begin{equation}
\label{e47}
\Hcal_{r} =  a\,r - {\alpha} \, \varsigma \,w \, \cos(\phi - \phi_0)-
{u}\,\varsigma\,\sin(\phi - \phi_0), 
\end{equation}
which coincides with $\Hcal_{r0}$ for  $\alpha =0$ ($a =1-\frac{\alpha}{2\, \pi} = 1$).\\ \vspace{-10pt}

 In both cases the world shells $\Hcal_{r_0} $ and $\Hcal_{r} $ are not invariant along the $r$-forms. Again, this can be cured by the addition of attached dislocations. For example, the edge dislocation along the $\{\textbf e_3 \}$ axis with Burgers vector $b_x = b\, \cos \phi_0, \,b_y = b\, \sin \phi_0,$:
 \begin{equation}
\label{e48}
ds^2 = -dt^2+(dx+ \cos \phi_0 \,\frac{b}{2\, \pi}\,d\phi)^2 +(dy+\sin \phi_0 \,\frac{b}{2\, \pi}\,d\phi)^2 +dz^2. 
\end{equation}  is singular on the hypersurface
\begin{equation}
\label{e49}
\Hcal_{e_{b}} =  r - \frac{b}{2\, \pi} \, \sin(\phi - \phi_0).  
\end{equation} Therefore dislocation densities $db = \sigma \,du$ attached to the $r$-form have the expected behavior. \\ \vspace{-10pt}
    
 \textsf{2- Another mixed $r$-form}\label{appli3} 

Equations \ref{e14} and~\ref{e45} both describe a disclination of angle $\alpha$ about the sheet $x^1 =x^2 = 0$ in the first case, $\lambda^1 =\lambda^2 = 0$ in the second; they coincide in the exchange $x^0 \leftrightarrow \lambda^0, x^i \leftrightarrow \lambda^i$. Consequently one can use the model of a $m$-twist disclination Eq.~\ref{e17} to construct a $r$-twist disclination. One gets:
\begin{equation}
\label{e50}
ds^2 = -(d\lambda^0)^2+(d\lambda^1+ \frac{\alpha}{2\, \pi}\,\lambda^3\,d\phi)^2 +(d\lambda^2)^2 +(d\lambda^3- \frac{\alpha}{2\, \pi}\,\lambda^1\,d\phi)^2.
\end{equation}\vspace{-35pt}

\subsection{Observability of the cut hypersurface of a cosmic form}\label{obscut}
  \vspace{-10pt}

This section concerns the $M^4$-crystals of Sect.~\ref{stab-cryst}, i.e., cases where the 4-momenta acquire a physical status.

As an example, let us consider a $m$-form, VP-constructed on a cut hypersurface $\Sigma_m$, defined by the symmetry element $A_m=\scriptsize{\begin{array}{|cc|} a & b\\  -\bar{b} & \bar{a} \end{array}}.$ $A_m$ leaves invariant any timelike 4-momentum as 
\begin{center}
$p_m^a = \{p_m^0,0,0,p_m^3\}, \quad \forall p_m^0, \, \forall p_m^3, \quad ||p_m^a||^2<0,$
\end{center}
and the resulting $m$-form is a legitimate VP defect in a spacetime inhabited only by the just defined 4-momenta. Call it a \textit{perfect} form. For any other 4-momentum field, the $m$-form shows up some imperfection: it differs from a perfect form by the appearance of a \textit{misorientation} from one side of $\Sigma_m$ to the other. Compare indeed $p$ on one side to $p' = A_m \,p\, A_m^*$ on the other for a 4-momentum different of the $p_m^a$'s just defined, e.g., $p_r^a=\{1,1,0,0\}$, say; one gets $ p'_r=A_m \,p_r\, A_m^* =\{1,\cos \alpha,\sin \alpha,0\}$. $\Sigma_m$ turns singular and made visible by this misorientation; the observability of the cut surface in space gives evidence of an \textit{imperfect} form. The cut hypersurface gets a physical status, similar to that one of a grain boundary in CMP. 
Momenta $p_m^a = \{p^0_m,0,0,p^3_m \}$ as above are continuous through the cut surface, but not on the world sheet or the world shells, which one expects to be the end or the start of incomplete geodesics. Since in that case $p_m^a$ does not suffer any singularity on the cut surface, the cosmic form is not sensitive to the exact location of $\Sigma_m$. 

These arguments extend easily to any cosmic form carrying a symmetry element $A_t$ or $A_r$. Thus the cut hypersurface of any $t$- or $r$-form is made physically observable in a spacetime which is not pCP, and thereby the cut surface is observable in any spacelike slice. \vspace{-20pt}

\subsection{Volterra defects in a de Sitter spacetime}\label{forms-dislo}
  \vspace{-10pt}

A dislocation is characterized by its Burgers vector $b^a \in \mathbb R^{1,3}$; $\mathbb R^{1,3}$ is nothing else than $M^4$ considered as a vector space. Thus we have timelike, null, and spacelike dislocations, as we have $m$-, $r$-, and $t$-disclination forms.  This is not a coincidence. Consider a de Sitter spacetime ${dS}_4$; its group of isometries is the indefinite orthogonal group $SO(1,4)$, which is also the Lorentz group for a 5D Minkowski space $M^5$. But in both cases the isotropy group at an event $x$, i.e. the subgroup which leaves $x$ invariant, is the Lorentz group $SO(1,3)$ (we have denoted all along this work by $L_0$ the connected part of $SO(1,3)$). Therefore both spaces appear as the quotients of their group of isometries by the same isotropy group:
\begin{equation}
\label{e51}
\mathbb{R}^{1,3}=M^4 =P(4)/SO(1,3), \qquad \qquad {dS}_4 = SO(1,4)/SO(1,3).
\end{equation}
\indent \textcite{inonu53} have shown in which sense a Lie group can be \textit{contracted} (this is their terminology) with respect to any of its continuous subgroups $S$, in such a way that the resulting group forms an abelian invariant subgroup of the contracted group. We quote:\textit{
 The subgroup $S$ with respect to which the contraction was undertaken
is isomorphic with the factor group of this invariant subgroup. Conversely,
the existence of an abelian invariant subgroup and the possibility to
choose from each of its cosets an element so that these form a subgroup $S$, is a necessary condition for the possibility to obtain the group from another group
by contraction.} Here, $SO(1,4)$ is contracted with respect to any of its subgroups $SO(1,3)$, which in this process tends towards $\mathbb{R}^{1,3}$, an invariant abelian subgroup of the Poincar\'e group: the disclinations turn into dislocations. Physically this contraction is equivalent to the increase without limit of the radius of curvature of $dS_4$, which turns it into $M^4$. 
 
 These remarks justify the use of the previous results on the classification of disclination cosmic forms for any maximally symmetric Lorentzian spacetime, but also the fact that a dislocation can be attached to a disclination. For this is nothing else than the result of a process of contraction from a virtual Lorentzian manifold with a non-vanishing curvature. \vspace{-10pt}
 
\section{Summary and discussion} \label{disc}
 \vspace{-10pt}

\subsection{Summing up the main characteristics of cosmic forms}\label{forms-main}
  \vspace{-10pt}

 A large part of the present article is devoted to the description of the three types of cosmic defects, the $m$-, $r$- and $t$-forms $-$ especially their metrics $-$ that originate in an analysis of any maximally symmetric Lorentzian manifold (like the Poincar\'e group or the de Sitter group) and of its subgroups in the spirit of the topological theory of defects. The defects relate to the classes of conjugacy of the subgroups of the Lorentz group, which themselves are partitioned into four sets, whether one considers timelike, null, or spacelike 4-momenta, or the momentum with vanishing components which is a class of conjugacy in itself. We have restrained to 2D defects (in a 4D spacetime), so that the defects we are interested in can be described as resulting from a Volterra process, a generalization of the cut-and-glue process classically employed for the creation of global cosmic strings: these defects carry a continuous invariant that is a continuous element of the Lorentz group. We have summarized in the introduction the main characteristics resulting from this group-theoretic point of view. Our discussion has been worked out all over for defects in a Minkowski spacetime $M^4$, but the results apply to a de Sitter spacetime, in which case the consideration of dislocations becomes needless. 
 
 We stress that cosmic forms are not resulting from a Kibble's mechanism at a phase transition. This mechanism yields quantized cosmic strings; cosmic forms are not quantized.
 
 To summarize the main results that emerge from our presentation of the three types of cosmic forms:
 
\indent \indent $-$ the Volterra process is the result of a relative displacement $-$ an element $g$ of the group of isometries of the spacetime $-$ of the lips of a 3D manifold $\Sigma$ (the cut hypersurface) bordered by a 2D manifold $\partial \Sigma$ (the world sheet); one differentiates wedge, twist, and mixed forms, according to the effect of $g$ on the world sheet, whether it leaves $\partial \Sigma$ invariant or not,

 \indent \indent $-$ \textit{wedge} cosmic $m$-, $t$- and $r$-forms are forms whose world sheets are invariant (locally or globally) under the action of $g$. The world sheet is then a 2-plane. Cosmic strings are analogous to {wedge} $m$-forms. 
  
  \indent \indent $-$ a \textit{twist} $m$-form is such that the world sheet is a 2-plane orthogonal to the 2D rotation axis; this world sheet is not invariant under tha action of $g$. The core of a twist $m$-form is hollow, the world sheet is non-singular; the singularity is carried by a 3D timelike manifold, the world shell whose size varies with the cosmic time.  A \textit{mixed} $m$-form is a generic non-wedge form, a twist form being a particular case,
     
   \indent \indent $-$ the world shell of a twist or mixed $m$-form can be stabilized at a constant size by the addition of a density of cosmic forms (dislocations or disclinations) attached all along the form; this property is analogous to the attachment of defect densities along a twist or a closed disclination loop (which necessarily includes twist components) in CMP \cite{klfr08},
   
      \indent \indent $-$ thus, a cosmic $m$-form carries any $\alpha$ and can have any 2D shape in 4D, provided its spacelike section is a closed loop; this is in contrast with global cosmic $m$-strings, which are usually thought of as 2-planes along the 2D rotation axis with quantized angle 
$\alpha \in SO(3) \subset L_0$. 

     \indent \indent $-$ the same analysis applies to $t$- and $r$-forms, but with differences originating in the more subtle nature of the rotation angles, which are now hyperbolic and null angles, respectively,
     
      \indent \indent $-$ the attachment process of infinitesimal defects does not depend whether the main disclination is of continuous or discrete strength; thus it applies to  cosmic strings. The dynamics through space of a string or a form certainly requires some change of shape of the line defect (some sort of fluctuation), thereby it is attended by variable defect densities, 
     
      \indent \indent $-$ in all the examples presented in this article, the calculation shows that the Riemann curvature of an isolated wedge or mixed form (before stabilization by attached densities) is vanishing (in $M^4$). We conjecture that this is always the case. But in the presence of attached densities the Riemann curvature is distorted around a mixed form. In fact, one expects that mixed forms are generically stabilized by attached forms, and thereby their Riemann curvature generically does not vanish,
     
      \indent \indent $-$ the defects we are concerned with act on some substance inhabiting the spacetime, for example a set of particles defined by a constant 4-momentum $p^a$. Let $H_a \subset L_0$ the group under which this set is invariant. The cut hypersurface $\Sigma_s$ of a cosmic form VP-constructed  from an element $A^s \in H_a$ is not observable because $p^a$ does not suffer any misorientation from one side of $\Sigma_s$ to the other.  Such a cosmic form is called \textit{perfect}. A cosmic form is \textit{imperfect} when its symmetry element $A_u$ acts on a constant field $p^b$ which is not invariant under the action of $A_u$, $A_u \notin H_b.$ The cut hypersurface $\Sigma_u$ is made visible.
   \vspace{-10pt}

\subsection{Physical properties, conjugated networks: a few conjectural remarks}\label{forms-net}
  \vspace{-10pt}
  
   \indent \indent $-$ The perfect cosmological principle. 
  As already mentioned, the $m$-, $t$- and $r$-forms belong to the Poincar\'e group or the indefinite orthogonal group $SO(1,4)$, according to the embedding spacetime $M^4$ or $dS_4$), etc. and as such obey the perfect cosmological principle (pCP) $-$ $t$- and $r$-forms are not compatible with the narrow cosmological principle (nCP).

 The steady state spacetime of Bondi et al. and Hoyle \cite{bondi48,*hoyle48}, namely the de Sitter spacetime $dS_4$, is built precisely in order to obey  pCP: it satisfies the Einstein's equations for a positive cosmological constant, usually interpreted as a \textit{false vacuum}; thus $r$- and $t$-forms are forms typical of a false vacuum. They could be met, as well as $m$-forms, during the process of inflation \cite{linde05ht}, as long as the spacetime is $dS_4$; as the decay of the false vacuum proceeds in function of time $t$, the spacetime takes a generic FRW structure and the $r$- and $t$-forms are no longer VP compatible with this structure: they might acquire a status of imperfect forms, or smoothly disappear, the group element they carry being continuous. \\ \vspace{-10pt}
 
 \indent \indent $-$ {The hollow core of a twist or mixed form in a pCP environment.}
S. \textcite{coleman77a} has proposed that the decay of the false vacuum results from the nucleation of bubbles of true vacuum in a false vacuum matrix; this model, used under different modifications in inflationary theory  (\textcite{vilenkin06}), is reminiscent of the CMP classical model of nucleation and growth at a first order phase transition. 

The nucleation and growth process of the true vacuum can also be assigned to mixed cosmic forms, because these objects always have a \textit{hollow core} separated from the outer part by a timelike hypersurface. If the core is filled with the true vacuum, the energy of the $r$-form results from a balance of the positive energy of the world shell and the energy of the true vacuum in the core, \textit{smaller} than the false vacuum energy ; this balance depends on the disclination characteristics ($ \alpha, \,\sigma, \, \omega $), yielding possibly a form whose energy is negative with respect to the false vacuum background. It is unusual to get a disclination energy that is negative, but it can be so if the disclination is thermodynamically required for phase stability, a phase transition or a metastable state. There are many examples in CMP. \\ \vspace{-10pt}
      
 \indent \indent $-$ {The hollow core of a wedge $m$- form in a nCP environment}. As stressed Sect.~\ref{mforms}2 the total mass-energy attached to a wedge segment can be also distributed on a 3D timelike circular cylinder centered on the world sheet, with the total mass $\mu$ independent of the cylinder radius $R$ \cite{langer70}. One can expect that this world shell could deform and displace easily, up to a point where it would be stabilized by the presence of foreign mass clusters. These remarks make sense if the deficit angle $\alpha >0$, since the mass-energy has to be positive. This does not preclude the existence of wedge $m$-forms with $\alpha <0$, if the core region is inhabited by a vacuum of \textit{larger} energy density than the vacuum of the outer region; such a situation might occur for a defect nucleated in the true vacuum, thus after the end of inflation. 
\\  \vspace{-10pt}

 \indent \indent $-$ Positive and negative cosmic form networks.
 Wedge $m$-forms are reminiscent of T. Regge's \textit{bones} \cite{regge61}, which carry continuous deficit angles. Bone segments form a \textit{skeleton} that approximate spacetime curvature, with specific Bianchi relations at the \textit{joints}. Similarly, one recognizes in CMP the existence of disclination \textit{networks} (i.e., skeletons), with Kirchhoff relations at the \textit{nodes} (i.e., joints), but also of \textit{conjugated disclination networks}, which are believed to be present in amorphous systems \cite{kleman79,kleman83}: each network is made of disclination segments which all of them carry deficit angles of the same sign (they cannot annihilate mutually), with opposite signs for the two networks. As a result the curvatures cancel on the average; the amorphous material can thereby live in a flat Euclidean space (as it must), at the expense of internal stresses.
   
 We speculate that cosmic forms gather into networks, with a richer structure than Regge's skeletons since they may incorporate $r$- and $t$-forms. Such networks would play a role in the setting up of the space curvature. This curvature almost vanishes at the end of the period of inflation ($|\Omega_{tot}-1|<10^{-30})$, without the intervention of disclinations according to the present views; this might also happens before inflation or at its beginning, by the nucleation of conjugated networks that could achieve a decrease of the occupied volume hence a decrease of the local energy (at constant false vacuum energy density). Thus, the inflation process would act on a already flat space.
 

But in reason of the present partial knowledge of the early universe, we stop here the stream of our speculations.
\vspace{-10pt}

\section*{Acknowledgments} I thank Jacques Friedel, Tom Kibble and Alex Vilenkin for discussions.

This is IPGP contribution \# 3154.
\newpage
\small
\bibliography{biblio}

\end{document}